\documentclass{article}
\usepackage{url}
\usepackage[pdftex]{graphicx}
\usepackage{amsmath}
\usepackage{booktabs}
\usepackage{amsfonts}
\usepackage{float}
\usepackage{pdflscape} 

\usepackage{subcaption} 
\usepackage{natbib}
\usepackage{tabularx} 
\usepackage{pdflscape} 
\usepackage{multirow}
\usepackage{authblk}  

\title{MedSR-Impact: Transformer-Based Super-Resolution for Lung CT Segmentation, Radiomics, Classification, and Prognosis}

\author{
    \textbf{Marc Boubnovski Martell},
    \textbf{Kristofer Linton-Reid}, 
    \textbf{Mitchell Chen}, 
    \textbf{Sumeet Hindocha}, 
    \textbf{Benjamin Hunter}, 
    \textbf{Marco A. Calzado},
    \textbf{Richard Lee}, 
    \textbf{Joram M. Posma}, 
    \textbf{Eric O. Aboagye} 
}

\begin{document}

\maketitle

\begin{abstract}
High-resolution volumetric computed tomography (CT) is essential for accurate diagnosis and treatment planning in thoracic diseases; however, it is limited by radiation dose and hardware costs. We present the Transformer Volumetric Super-Resolution Network (\textbf{TVSRN-V2}), a transformer-based super-resolution (SR) framework designed for practical deployment in clinical lung CT analysis. Built from scalable components Through-Plane Attention Blocks (TAB) and Swin Transformer V2—our model effectively reconstructs fine anatomical details in low-dose CT volumes and integrates seamlessly with downstream analysis pipelines. 

We evaluate its effectiveness on three critical lung cancer tasks—lobe segmentation, radiomics, and prognosis—across multiple clinical cohorts. To enhance robustness across variable acquisition protocols, we introduce pseudo-low-resolution augmentation, simulating scanner diversity without requiring private data. TVSRN-V2 demonstrates a significant improvement in segmentation accuracy (+4\% Dice), higher radiomic feature reproducibility, and enhanced predictive performance (+0.06 C-index and AUC). These results indicate that SR-driven recovery of structural detail significantly enhances clinical decision support, positioning TVSR-V2 as a well-engineered, clinically viable system for dose-efficient imaging and quantitative analysis in real-world CT workflows.

\end{abstract}

\noindent\textsuperscript{*} Equal contribution. Listing order is random.

\section{Introduction}
\label{sec:intro}

High-resolution computed tomography (CT) is essential in the detection, diagnosis, and management of lung cancer, particularly non--small cell lung cancer (NSCLC). Thin-slice CT (typically $\leq 1.5$ mm) allows for the detailed visualization of pulmonary nodules, airway structures, and vascular anatomy, which is critical for early-stage lung cancer assessment \cite{macmahon2017guidelines, hattori2017prognostic}. Studies have shown that thin-slice CT improves the detection of subsolid nodules and assists in distinguishing between pre-invasive and invasive adenocarcinomas \cite{li2023thin}. These imaging advantages directly impact patient outcomes by enabling timely interventions \cite{martell2025radiomics}.

Despite its clinical utility, thin-slice CT is not routinely available in all healthcare settings. Acquiring high-resolution images requires either higher radiation doses, which may pose safety concerns \cite{boas2012ct}, or expensive scanning and storage infrastructure \cite{park2021performance}. Many clinical protocols default to thick-slice reconstruction (e.g., 5 mm) to manage resource limitations \cite{zhu2025designing}. This compromises spatial resolution and diagnostic sensitivity—particularly for small or subtle lesions—and poses a challenge to AI tools trained predominantly on high-resolution data \cite{he2016effects, gupta2023impact}.

To address these limitations, super-resolution (SR) techniques have emerged as a promising solution. SR aims to reconstruct high-resolution images from low-resolution inputs using data-driven mappings. Deep learning–based SR models, particularly convolutional neural networks (CNNs), have shown substantial improvements over traditional interpolation methods in recovering anatomical details \cite{dong2015image, zhang2018residual}. However, most SR models are developed for 2D natural images and fail to generalize well to 3D medical data such as volumetric CT, where through-plane consistency is crucial for anatomical fidelity.

Recent advances in Transformer-based architectures—such as SwinIR \cite{liang2021swinir}, HAT \cite{chen2023hat}, and ART \cite{zhang2023accurate}—have demonstrated state-of-the-art results in 2D SR by modeling long-range dependencies with self-attention. In the medical domain, transformer models are beginning to be applied to 3D imaging tasks, including volumetric SR for CT and MRI \cite{yu2022rplhr}. These approaches have shown promise in improving spatial resolution while maintaining global structural integrity.

In this work, we introduce \textbf{TVSRN-V2}, a transformer-based volumetric CT SR model specifically designed to recover high-resolution anatomical structures from thick-slice CT scans. TVSRN-V2 integrates Swin Transformer V2 layers and novel Through-Plane Attention Blocks (TAB) within an asymmetric encoder–decoder architecture to capture inter-slice dependencies. Unlike conventional SR models evaluated solely on image fidelity metrics and similar to recent work \cite{yu2024spatial}, we assess TVSRN-V2’s impact on downstream clinical tasks, including lung segmentation, radiomic reproducibility, histology classification, and prognosis in NSCLC patients.

\textbf{Our key contributions are:}
\begin{itemize}
  \item We present TVSRN-V2, a transformer-based SR model optimized for volumetric CT with explicit modeling of through-plane anatomical consistency.
  \item We propose a hybrid training strategy combining real and pseudo low-resolution CT scans to improve generalization across variable slice thicknesses.
  \item We validate the clinical relevance of SR by evaluating its impact on segmentation accuracy, radiomic feature reproducibility, and downstream predictive modeling in NSCLC.
\end{itemize}

\section{Background}

\subsection{CT Resolution in Lung Cancer}

Slice thickness in CT imaging is a key determinant of spatial resolution and diagnostic accuracy. Thin-slice CT ($\leq 1.5$ mm, typically 1mm) enables precise visualization of pulmonary structures, facilitating early detection and characterization of small nodules \cite{wang2022different}. Clinical guidelines recommend thin-slice reconstructions for evaluating incidental lung nodules and suspected malignancies \cite{macmahon2017guidelines}.

Despite their diagnostic advantages, thin-slice CT scans are not consistently available in clinical practice due to higher radiation dose, storage demands, and reconstruction complexity. Many scanners can acquire thin slices, but protocols often default to thick-slice (5mm) settings due to infrastructure limitations \cite{boas2012ct, park2021performance}. This issue is especially acute in low-resource settings, where thick-slice CT remains standard and may obscure subtle anatomical features critical for lung cancer diagnosis \cite{gupta2023impact}.

Thick-slice CT also presents challenges for AI-based tools. Most deep learning models are developed and validated on high-resolution datasets and show reduced performance on coarse-resolution inputs, exacerbating disparities in diagnostic access \cite{he2016effects}. Bridging this resolution gap is essential to enable robust AI-assisted diagnosis across diverse clinical environments.

More recently, several studies have demonstrated the benefits of SR techniques on downstream tasks using CT imaging \cite{umirzakova2024medical}. These advancements can enhance feature robustness by standardizing features across multiple cohorts \cite{de2021impact}, thereby reducing segmentation time and improving overall image quality \cite{abbott2025super}. Additionally, a study demonstrated that a deep learning model can effectively generate high-quality synthetic thin-slice CT from thick-slice CT, with diagnostic accuracy for pneumonia and lung nodules comparable to real thin-slice CT \cite{yu2024spatial}.

\subsection{SR Techniques for CT}

SR aims to reconstruct high-resolution (HR) images from low-resolution (LR) inputs and is particularly relevant in CT imaging, where slice thickness significantly impacts diagnostic precision. Traditional interpolation methods are limited in restoring anatomical detail, while deep learning-based SR has shown promise in enhancing volumetric image quality.

Three primary deep learning paradigms are used for SR: Generative Adversarial Networks (GANs), Transformers, and Diffusion Models. GAN-based approaches \cite{you2019ct, wang2018esrgan} generate perceptually realistic images via adversarial learning but may produce artifacts and are difficult to train. Diffusion models \cite{gao2023implicit,wang2024sinsr} iteratively refine images through noise modeling and offer strong reconstruction capabilities, though they are computationally intensive. Transformer-based methods such as SwinIR \cite{liang2021swinir}, HAT \cite{chen2023hat}, and ART \cite{zhang2023accurate} leverage global self-attention to capture long-range dependencies, setting new benchmarks on standard SR datasets.

Notably, most SR models are developed for 2D natural images and do not generalize well to 3D medical volumes. In CT imaging, through-plane consistency is critical for diagnostic integrity but is often overlooked. Early CNN-based models such as SRCNN \cite{dong2015image} and RDN \cite{zhang2018residual} were adapted for SR in CT but were limited in their ability to model spatial context. Later attention-based models like RCAN \cite{zhang2018image} improved channel-wise feature learning, and SwinIR introduced shifted-window attention, though its use remained primarily 2D.

To address these limitations, we propose TVSRN-V2 tailored to volumetric images, a transformer-based architecture tailored for volumetric CT SR. Building on prior work \cite{yu2022rplhr}, our model incorporates Swin Transformer V2 \cite{liu2022swin} and Through-Plane Attention Blocks (TAB) to model inter-slice dependencies effectively. Unlike existing approaches that focus solely on pixel-level metrics, we evaluate TVSRN-V2 on downstream tasks including segmentation, radiomics, and prognosis to demonstrate its clinical utility.

\section{Methods}

\subsection{Dataset}\label{subsec:dataset}
\subsubsection{SR Dataset}

We used the publicly available RPLHR-CT dataset \cite{yu2022rplhr}, which contains 250 anonymized clinical CT scans in NIfTI format. All scans were acquired on Philips scanners and reconstructed into two resolutions: thin-slice (1 mm) and thick-slice (5 mm). The number of slices varied from $L\in[191, 396]$ for thin CT and $L\in[39, 80]$ for thick CT, with consistent in-plane resolution across both ($0.604$–$0.795$ mm). All scans are aligned using patient coordinate systems.

\begin{figure}[h]
    \centering
    \includegraphics[width=0.95\linewidth]{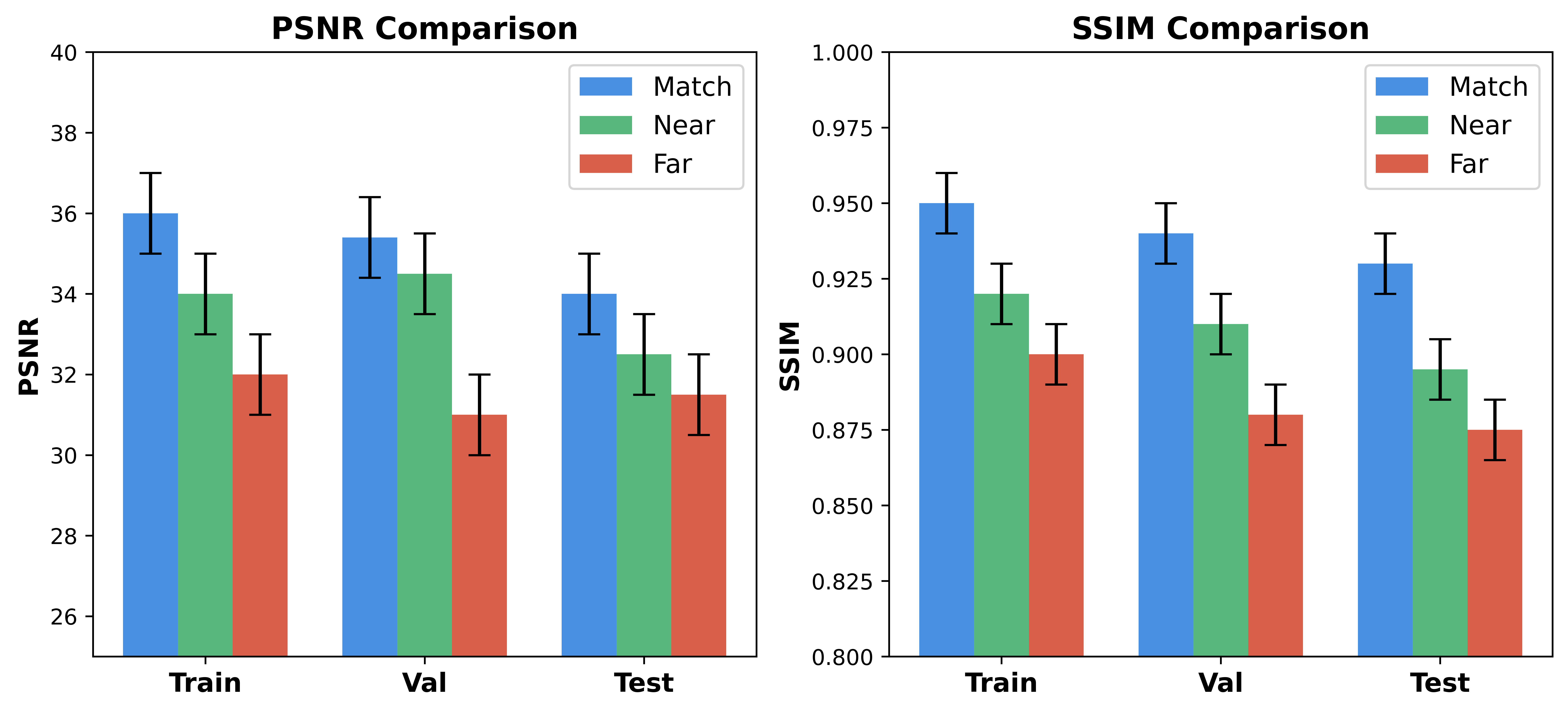}
    \caption{\textbf{Slice similarity across spatial distances.} PSNR and SSIM comparisons for slice-pairs with varying spatial offsets with match, near and far with respective distances (0 mm, 1 mm, 2 mm).}
    \label{fig:SimilarityPerSlice}
\end{figure}

We split the dataset into 100 training, 50 validation, and 100 test volumes. To assess spatial correspondence between thin and thick slices, we grouped slice-pairs by distance and compared them using PSNR and SSIM metrics (Figure~\ref{fig:SimilarityPerSlice}). Slices at the same spatial location showed the highest similarity, which decreased with increasing slice separation.

To increase training diversity, we applied slice-wise downsampling to the HR CTs along the through-plane axis to create pseudo low-resolution volumes. Downsampling was performed until either the slice thickness exceeded 3 mm or the total number of slices fell below 130. This augmentation strategy follows Peng et al. \cite{peng2020saint} but avoids excessive degradation by capping at 3 mm, as thicker pseudo-CTs are less effective for training \cite{yu2022rplhr}. Fine-tuning was done using both the real and pseudo low-resolution scans, enabling a broader spectrum of slice thicknesses while preserving anatomical consistency. This setup better reflects real-world conditions, where CT slice spacing varies between 0.5 and 6 mm \cite{huang2021impact}.

\subsubsection{Datasets for Evaluating the Impact of SR on Feature Robustness}\label{sec:RIDER}

To evaluate the consistency of the TMR-CT features, a test-retest experiment was performed
using the RIDER dataset, which included 32 patients with lung cancer \cite{sung2021global}.

\subsubsection{Datasets for Evaluating the Impact of SR on  Lung Segmentation}\label{subsec:data_seg_summary}

To train and evaluate the effect of SR preprocessing on segmentation, we used diverse CT datasets covering both normal and pathological lung conditions.

\textbf{Training set.} We used 50 thoracic CT scans with manual segmentations of lobes, bronchi, and trachea, validated by a board-certified radiologist. These included 38 publicly available scans from the SPIE-AAPM Lung CT Challenge \cite{armato2016lungx} and 12 local scans from Imperial College Healthcare NHS Trust (REC: 18HH4616) \cite{boubnovski2022development}.

\textbf{Validation/test sets.} 50 scans from the LUNA16 dataset \cite{tang2019automatic}, derived from the LIDC-IDRI cohort \cite{armato2011lung}, were used. The dataset includes CTs acquired on different scanners with varying slice thicknesses ($\leq$3mm). Half were used for validation and half for testing.

\textbf{External pathological cohort.} We additionally tested generalization on 24 patients across four disease categories: COPD, lung cancer, COVID-19 pneumonitis, and lobar collapse. These were sourced from local and public datasets \cite{boubnovski2022development, gevenois1995comparison}, with emphysema quantified via low-attenuation areas and density percentiles.

\subsubsection{Datasets for Evaluating the Impact of SR on Classification and Prognosis}\label{sec:prognosis}

We evaluate the effect of SR on two clinically relevant downstream tasks—histology classification and survival prediction of NSCLC—using a multi-institutional dataset spanning four centers. The training was performed on the public TCIA cohort (n=203), while external validation was conducted on three UK hospitals from the OCTAPUS-AI study: GSTT, ICHT, and RMH (n=539 total) \cite{hindocha2022comparison}. The datasets vary in slice thickness, CT contrast settings, demographics, and treatment patterns. This diversity allows us to assess the generalizability of SR-enhanced features across real-world clinical variation. 

{Further cohort statistics (e.g., stage, dose, gender distribution)} are provided in the supplementary material (Table \ref{tab:institution_comparison}).

\subsection{TVSRN-V2: Volumetric CT SR with Swin Transformer V2}
TVSRN-V2 is an asymmetric encoder-decoder architecture designed for volumetric CT SR. It reconstructs missing slices in low-resolution scans using contextual information from surrounding slices. Compared to the original TVSRN \cite{yu2022rplhr}, TVSRN-V2 integrates Swin Transformer V2 \cite{liu2022swin} for more scalable and stable training while enabling higher model capacity on standard GPUs.

\begin{figure*}[htbp]
  \centering
  \includegraphics[height=0.3\textheight,width=0.8\textwidth]{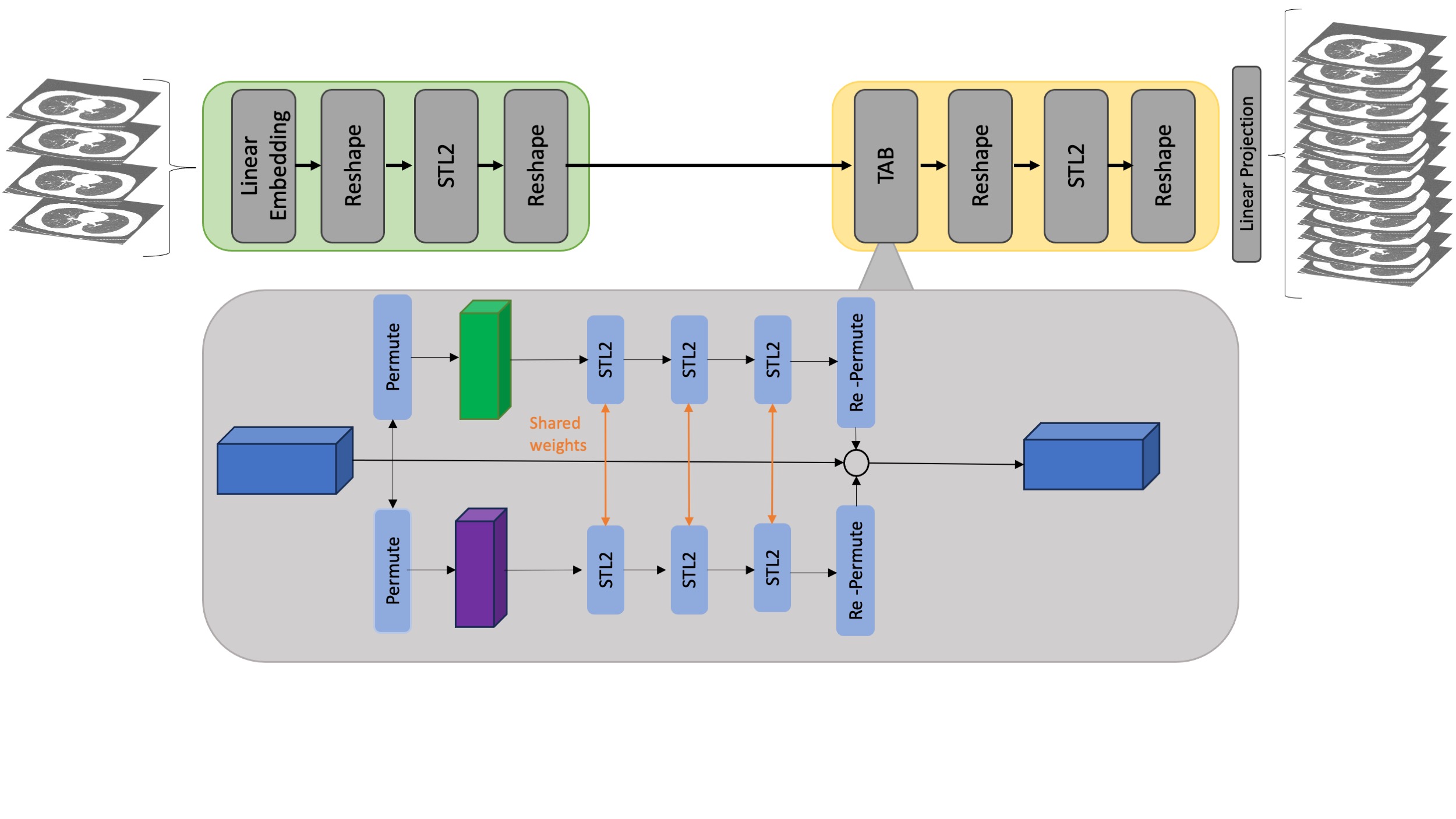}
  \caption{\textbf{TVSRN-V2 architecture.} The encoder (green) extracts low-resolution features; the decoder (yellow) reconstructs the high-resolution slices using Through-Plane Attention Blocks (TABs).}
  \label{fig:TVSRN-V2}
\end{figure*}

As shown in Figure \ref{fig:TVSRN-V2}, the encoder extracts features from the low-resolution input through linear embedding and Swin Transformer Layers (STL2). The decoder, comprising Feature Interaction Modules (FIMs), reconstructs the full-resolution volume using Through-Plane Attention Blocks (TABs) and STL2 blocks.

TAB processes an input volume \(X \in \mathbb{R}^{C \times D \times H \times W}\) through dual attention pathways to capture comprehensive volumetric context. Given a latent tensor \(z_{\text{in}}\), the attention mechanism operates as:
\begin{equation}
\begin{aligned}
   z_0^{\text{sag}} &= P^{\text{sag}}(z_{\text{in}}), \quad z_0^{\text{cor}} = P^{\text{cor}}(z_{\text{in}}), \\
   z_j^{\text{sag}} &= H_j^{\text{STL}}(z_{j-1}^{\text{sag}}), \quad z_j^{\text{cor}} = H_j^{\text{STL}}(z_{j-1}^{\text{cor}}), \\
   z_{\text{out}} &= z_{\text{in}} + P_{\text{re}}^{\text{sag}}(z_4^{\text{sag}}) + P_{\text{re}}^{\text{cor}}(z_4^{\text{cor}}).
\end{aligned}
\label{eq:TAB}
\end{equation}

The TAB first projects features into query/key/value tensors and processes them through two complementary paths: (1) Sagittal attention (\(z^{\text{sag}}\)) operating on depth D=64 as sequence length, enabling slice-wise feature correlation, and (2) Coronal attention (\(z^{\text{cor}}\)) processing height H=256 to capture in-plane anatomical relationships. The permutation operators \(P\) and \(P_{\text{re}}\) enable efficient view transformations, while \(H_j^{\text{STL}}\) applies Swin V2 attention within each view. This dual-pathway design effectively models both through-plane and in-plane dependencies with minimal computational overhead.

\begin{figure}[htbp]  
  \centering
  \begin{subfigure}[b]{0.45\textwidth}  
    \includegraphics[width=\textwidth]{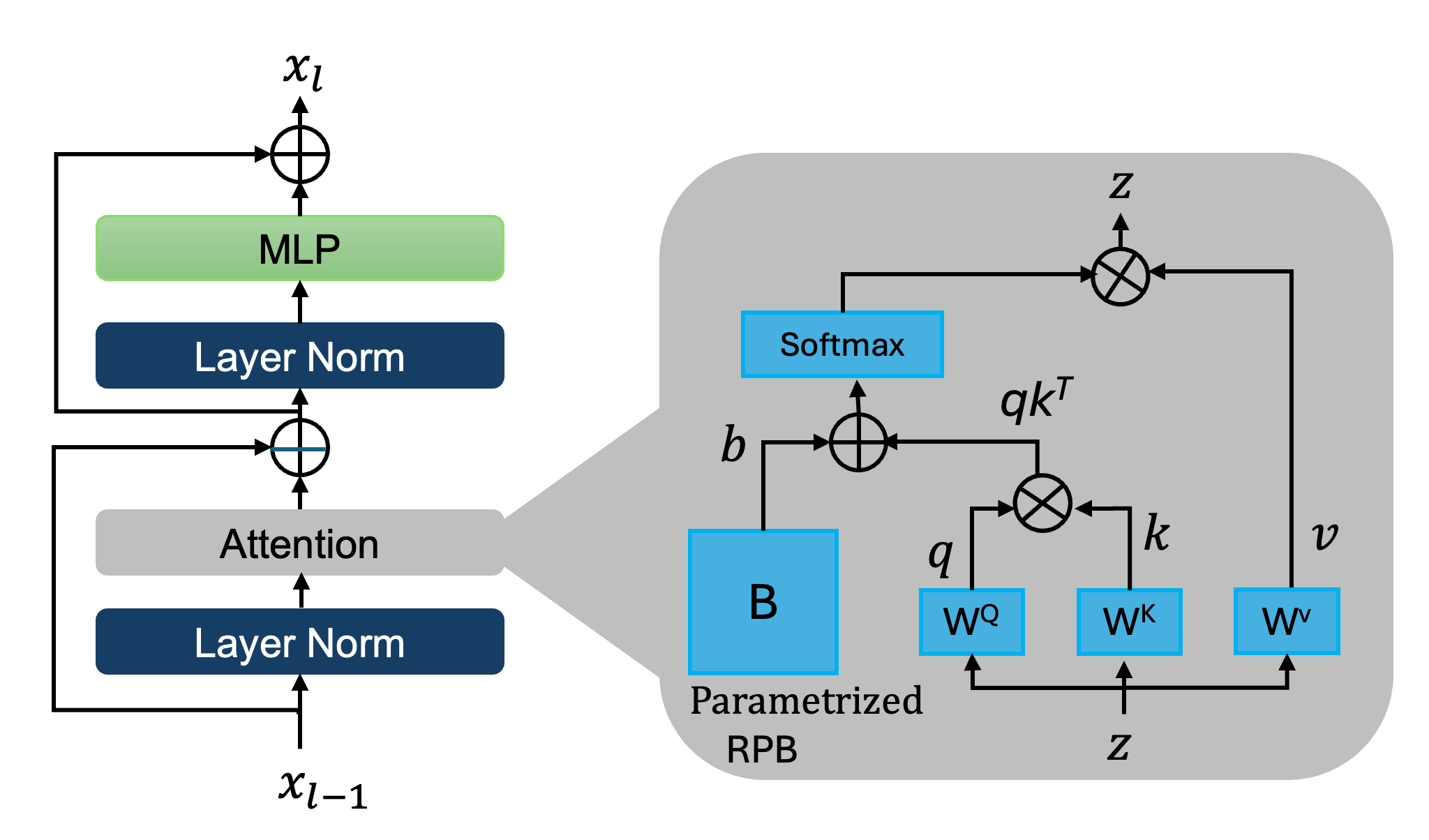}  
    \caption{Swin V1}  
    \label{fig:swin_v1}
  \end{subfigure}  
  \hspace{0.02\textwidth}  
  \begin{subfigure}[b]{0.45\textwidth}  
    \includegraphics[width=\textwidth]{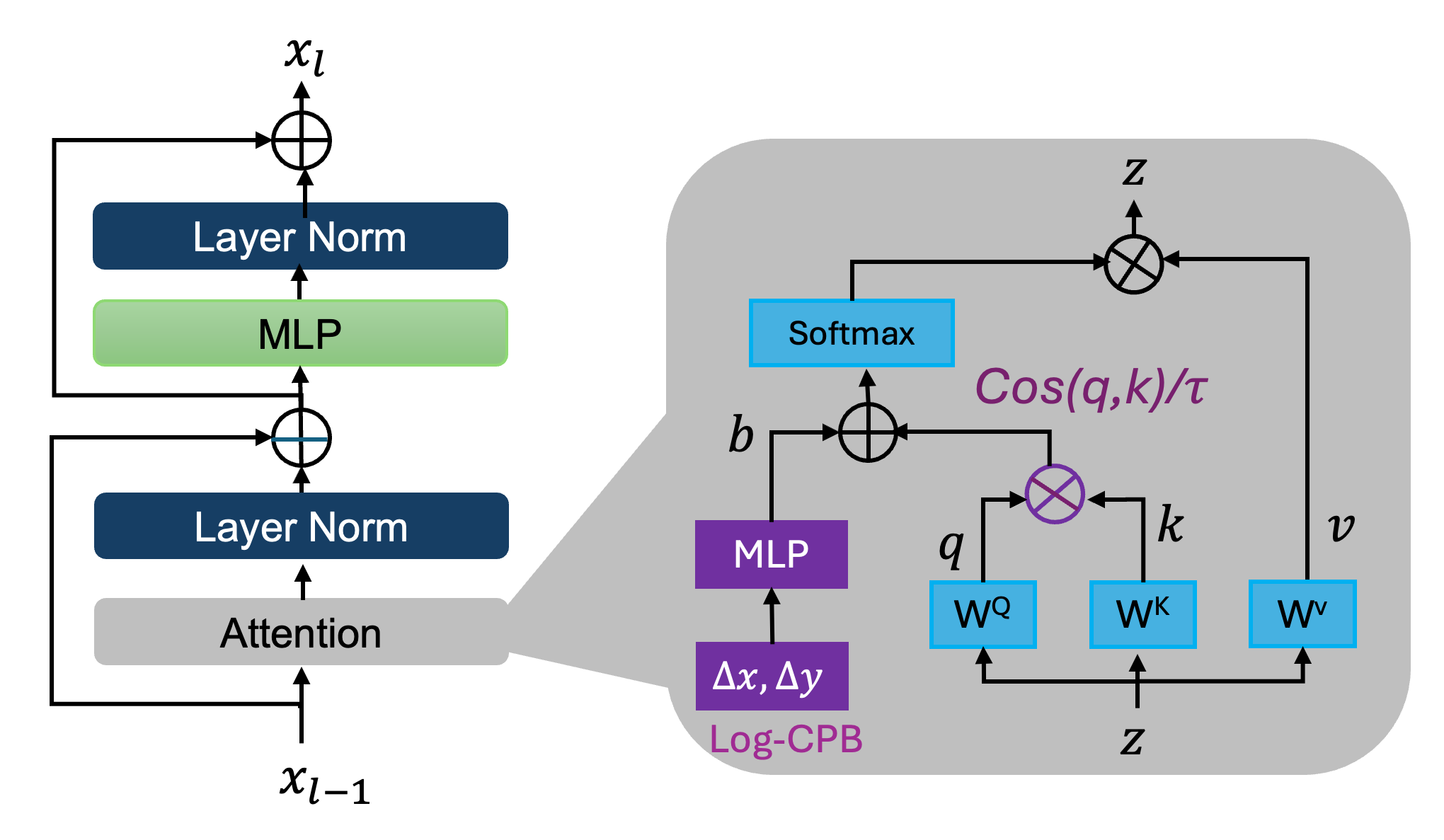}  
    \caption{Swin Transformer V2}
    \label{fig:ig:swin_v2}
  \end{subfigure}  
  \caption{\textbf{Swin Transformer V1 vs. V2.} Swin Transformer V2 introduces scaled cosine attention, residual post-normalization, and continuous log-space relative positional bias.}
  \label{fig:swinv1_v_v2}  
\end{figure}

TVSRN-V2 incorporates key Swin V2 improvements (Figure \ref{fig:swinv1_v_v2}): continuous relative positional bias via MLP-generated bias, residual post-normalization, and scaled cosine attention:
\begin{equation}
\mathrm{Sim}(\mathbf{q}_i, \mathbf{k}_j) = \frac{\cos(\mathbf{q}_i, \mathbf{k}_j)}{\tau} + B_{ij},
\label{eq:new_att}
\end{equation}
where \(\tau\) is a learnable temperature parameter and \(B_{ij}\) is the relative positional bias.

The model is trained using an \(L_1\) loss:
\begin{equation}
L_{\text{$_1$}} = \frac{1}{D' \times H \times W} \sum_{d,h,w} \left| \hat{Y}_{d,h,w} - Y_{d,h,w} \right|,
\label{eq:l1_loss_res}
\end{equation}
where \(\hat{Y}\) and \(Y\) are the predicted and ground truth volumes. Training employs activation checkpointing and ZeRO optimization \cite{chen2016training, rajbhandari2020zero} for efficient resource utilization.

\subsection{Implementation Details}

TVSRN-V2 was implemented in PyTorch and trained on NVIDIA A6000 GPUs. The model was trained with a batch size of 1 using randomly sampled input cubes of size $4{\times}256{\times}256$ from thin-slice CT scans. The corresponding targets were $16{\times}256{\times}256$ volumes. CT intensity values were clipped to $[-1024, 2048]$ and normalized to $[0, 1]$. Training used the Adam optimizer with a learning rate of $1\mathrm{e}^{-4}$. Data augmentation included random cropping and horizontal flipping.

Initial training was performed exclusively on real low/high-resolution CT pairs using 15 patients over 2000 epochs. Fine-tuning was subsequently performed using a mix of real and pseudo low-resolution scans.

At inference, input cubes of size $4{\times}256{\times}256$ were extracted from thick-slice CTs using a sliding window with one-slice overlap along the depth axis. For overlapping predictions, voxel-wise averaging was applied. If fewer than four slices were available, the last slices were repeated to maintain the required input depth.

\subsection{Evaluation Protocols for Downstream Tasks}
\label{subsec:evaluation_protocols}

To assess the impact of super-resolution (SR) on downstream medical imaging tasks, we designed five evaluation protocols spanning segmentation, radiomic reproducibility, and prognosis/classification performance. These protocols enable a systematic investigation of SR's utility in real-world low-resolution CT scenarios where ground truth segmentations or high-resolution acquisitions may be limited or unavailable.

\subsubsection{Radiomic and TMR-CT Feature Reproducibility}
We analyzed the stability of radiomic and TMR-CT features using a modified version of the RIDER dataset (Section~\ref{sec:RIDER}), where each subject had two CT scans: one high-resolution and one intentionally downsampled. Radiomic features were extracted using methodologies described in the study by Lu et al. \cite{lu2019mathematical}, which presents a mathematical descriptor for tumor structures from CT images that correlate with prognostic and molecular phenotypes. Similarly, TMR-CT features were derived as outlined in Boubnovski et al. \cite{boubnovski2024deep}, where deep representation learning of tissue metabolomes and computed tomography informs NSCLC classification and prognosis. Features were extracted with and without SR enhancement using standardized software tools, and intra-subject reproducibility was assessed to determine whether SR mitigates resolution-induced feature drift.

\subsubsection{Segmentation on Real Low-Resolution CTs}
Using the test cohort described in Section~\ref{subsec:data_seg_summary}, we applied SR to CT scans with inter-slice spacing exceeding 1 mm. The SR-enhanced volumes were then used to train and evaluate a V-Net model with multi-task learning (MTL) tailored for lobar segmentation. Notably, this model segments smaller structures, such as lobar bronchi, which may significantly aid in overall lobe segmentation accuracy.

\subsubsection{Segmentation on Pseudo Paired CTs}
To create paired pseudo low/high-resolution datasets, we subsampled slices from 50 healthy and 24 diseased CT volumes as detailed in Section~\ref{subsec:data_seg_summary}. Lobar segmentation masks derived from the original high-resolution volumes were utilized as references. This setup allowed for controlled evaluation of whether SR improves segmentation accuracy under conditions with known degradation.

\subsubsection{Cross-Resolution Evaluation with Real CT Pairs}
We utilized 100 real paired low- and high-resolution scans (Section~\ref{subsec:data_seg_summary}) to simulate clinically realistic scenarios. A V-Net model trained on high-resolution CTs was applied to both HR scans and SR-enhanced LR scans. The high-resolution-derived masks served as proxy ground truth to evaluate whether SR can recover meaningful structural detail lost due to thick-slice imaging practices.

\subsubsection{Impact on Classification and Prognosis Models}
Finally, we evaluated whether SR improves performance in downstream predictive models \cite{xing2024deep,lancet2023end}. For histology classification and prognosis prediction (Section~\ref{sec:prognosis}), CT scans containing fewer than 200 slices were enhanced using TVSRN-V2. We then compared model outputs on SR-enhanced scans against those on native low-resolution scans to assess whether SR serves as an effective preprocessing step for clinical inference.

\section{Experiments and Results}
\subsection{Ablation}

An ablation study was conducted to identify key components that affect the performance of TVSRN-V2. Results across three distinct variations were reported to assess their significance. The variations include:

\begin{enumerate}
\item \textbf{TVSRN-V2$^{w/o TAB}$}: This variant of the TVSRN-V2 model excludes the Through-plane Attention Block (TAB), which neglects the relative positioning of the slices during the SR process, leading to reduced performance.
\item \textbf{TVSRN-V2$^{Encoder}$}: In this model, only the TVSRN-V2 encoder is utilized, with upsampling performed through a subpixel conversion method as proposed by Shi et al. \cite{shi2016real}. This variant demonstrates the limitations of the encoder alone, highlighting the impact of the full architecture.
\item \textbf{TVSRN-V2$_{ViT}^{Encoder}$}: This version employs a standard transformer-based model, as introduced by Dosovitskiy et al. \cite{dosovitskiy2020image}, and similarly uses the subpixel conversion method for upsampling, providing a baseline comparison with a conventional architecture.
\end{enumerate}

\begin{table}[h]
\centering
\scriptsize
\caption{Ablation study results for TVSRN-V2 performance on the internal test set $\pm$ standard deviation. An asterisk (*) indicates a statistically significant difference ($p<0.05$) defined by the one-sided Wilcoxon signed-rank test between the method and all other methods with worse performance. The best results are shown in bold.}
\label{table:denoise_ablation}
\begin{tabular}{llcc}
\toprule
Designs & PSNR(↑) & SSIM(↑) \\
\midrule
TVSRN-V2$^{w/o TAB}$  & 37.820 ± 1.834* & 0.915 ± 0.023 \\
TVSRN-V2$^{Encoder}$  & 36.456 ± 1.675* & 0.900 ± 0.026 \\
TVSRN-V2$_{ViT}^{Encoder}$  & 34.688 ± 1.278 & 0.871 ± 0.025 \\
TVSRN-V2  & \textbf{39.162} ± \textbf{1.834}* & \textbf{0.941} ± \textbf{0.028}* \\
\bottomrule
\end{tabular}
\end{table}

Table~\ref{table:denoise_ablation} shows the performance of TVSRN-V2 compared to its variations. The results indicate that the inclusion of the full architecture, particularly with the integration of the Through-plane Attention Block (TAB), is critical for achieving superior performance.

TVSRN-V2$^{w/o TAB}$ shows a decrease in both PSNR and SSIM, indicating that excluding the TAB significantly hampers the model's ability to capture slice relationships effectively. Furthermore, the performance of TVSRN-V2$^{Encoder}$ is markedly lower, demonstrating that relying solely on the encoder without the enhancements provided by the decoder and TAB limits the model's capabilities.

The substantial improvement observed in TVSRN-V2 suggests that incorporating relative positional relationships in slice processing, as well as the complete architecture, leads to enhanced performance, making it a highly effective and robust solution for volumetric CT super-resolution tasks.

  \subsection{SR Performance}

To evaluate TVSRN-V2, the model was compared against Bicubic interpolation as a baseline and the original TVSRN \cite{yu2022rplhr}. For a model to be considered effective, it must recreate high-resolution images while preserving and enhancing finer details during reconstruction. In the case of TVSRN-V2, it achieved an SSIM of \(0.946 \pm 0.028\) and a PSNR of \(39.162 \pm 1.834\), demonstrating significant improvements over the other models (\(p<0.05\)), as shown in Table \ref{table:results_denoise_internal_clean}. 

A high SSIM indicates that the inherent and visual structures of the images are well-preserved, reflecting strong alignment with the original content. Similarly, the elevated PSNR suggests that the reconstructed high-resolution images are cleaner and sharper, enhancing the overall visual quality.

\begin{table}[h]
\centering
\scriptsize  
\caption{Performance on the internal test set ($\pm$ std). 95\% confidence intervals are shown in brackets. Asterisk (*) indicates $p<0.05$ vs. TVSRN-V2 (Wilcoxon signed-rank test). Best results in bold.}
\label{table:results_denoise_internal_clean}
\begin{tabular}{lcc}
\toprule
Model & PSNR (↑) & SSIM (↑) \\
\midrule
Bicubic & 33.214 $\pm$ 1.083* & 0.904 $\pm$ 0.028* \\
& [31.628, 35.225] & [0.821, 0.932] \\
TVSRN & 38.578 $\pm$ 1.621* & 0.937 $\pm$ 0.021* \\
& [36.012, 41.202] & [0.896, 0.969] \\
TVSRN-V2 & \textbf{39.162} $\pm$ \textbf{1.834} & \textbf{0.946} $\pm$ \textbf{0.028} \\
& [\textbf{37.438}, \textbf{42.122}] & [\textbf{0.902}, \textbf{0.977}] \\
\bottomrule
\end{tabular}
\end{table}

The performance results underscore the efficacy of TVSRN-V2 in producing high-resolution images that retain fine structural details. The PSNR improvement over the original TVSRN model demonstrates that the incorporation of advanced architectural components, such as the Through-plane Attention Block (TAB) and Swin Transformer V2 layers, is crucial for achieving superior reconstruction quality. 
\subsection{Post-SR Lobar Segmentation}\label{subsec:SR_result_seg}
This section evaluates the effectiveness of TVSRN-V2 as a preprocessing step for lobar segmentation using the V-Net MTL model \cite{boubnovski2022development}. Following the application of various SR techniques, segmentation was performed on chest CT scans, and the results were compared using Dice scores.

We assess segmentation performance on: (1) standard-resolution CT scans, (2) pseudo low-resolution CT scans generated from the originals, and (3) real low-resolution CT scans. Each was enhanced with different SR methods before segmentation.

\begin{table}[t]
    \centering
    \scriptsize  
    \caption{Dice scores for V-Net MTL on normal-resolution CT with different SR preprocessing. Bold indicates best. * denotes $p<0.05$ vs. Original.}
    \label{tab:dice_scores_normal_singlecol}
    \begin{tabular}{lccc}
        \toprule
        Class & Original & TVSRN & TVSRN-V2 \\
        \midrule
        LR lobe & 0.962 $\pm$ 0.041 & 0.965 $\pm$ 0.038 & \textbf{0.967} $\pm$ \textbf{0.042} \\
        MR lobe & 0.931 $\pm$ 0.064 & 0.935 $\pm$ 0.059 & \textbf{0.938} $\pm$ \textbf{0.057} \\
        UR lobe & 0.952 $\pm$ 0.047 & 0.962 $\pm$ 0.055 & \textbf{0.964} $\pm$ \textbf{0.049} \\
        LL lobe & 0.967 $\pm$ 0.043 & 0.966 $\pm$ 0.048 & \textbf{0.970} $\pm$ \textbf{0.037} \\
        UL lobe & 0.969 $\pm$ 0.046 & 0.968 $\pm$ 0.047 & \textbf{0.972} $\pm$ \textbf{0.025}* \\
        Trachea$^{b}$ & 0.972 $\pm$ 0.052 & 0.979 $\pm$ 0.038 & \textbf{0.985} $\pm$ \textbf{0.023}* \\
        Bronchi$^{b}$ & 0.649 $\pm$ 0.123 & 0.719 $\pm$ 0.143 & \textbf{0.748} $\pm$ \textbf{0.129}* \\
        \bottomrule
    \end{tabular}
\end{table}

Table~\ref{tab:dice_scores_normal_singlecol} shows segmentation performance on standard-resolution scans. While all SR methods improved results, TVSRN-V2 consistently yields the highest Dice scores, particularly on small or complex structures such as bronchi and trachea.

\begin{table}[t]
    \centering
    \scriptsize  
    \caption{Dice scores for V-Net MTL on pseudo low-resolution CT after SR. Bold indicates best. * denotes $p<0.05$ vs. Bicubic.}
    \label{tab:dice_scores_pseudo_singlecol}
    \begin{tabular}{lccc}
        \toprule
        Class & Bicubic & TVSRN & TVSRN-V2 \\
        \midrule
        LR lobe & 0.917 $\pm$ 0.033 & 0.949 $\pm$ 0.034* & \textbf{0.955} $\pm$ \textbf{0.030}* \\
        MR lobe & 0.899 $\pm$ 0.032 & 0.922 $\pm$ 0.028* & \textbf{0.933} $\pm$ \textbf{0.027}* \\
        UR lobe & 0.913 $\pm$ 0.029 & 0.951 $\pm$ 0.035* & \textbf{0.956} $\pm$ \textbf{0.028}* \\
        LL lobe & 0.907 $\pm$ 0.025 & 0.941 $\pm$ 0.028* & \textbf{0.952} $\pm$ \textbf{0.024}* \\
        UL lobe & 0.909 $\pm$ 0.022 & 0.944 $\pm$ 0.031* & \textbf{0.960} $\pm$ \textbf{0.027}* \\
        Trachea$^{b}$ & 0.922 $\pm$ 0.031 & 0.962 $\pm$ 0.035* & \textbf{0.968} $\pm$ \textbf{0.028}* \\
        Bronchi$^{b}$ & 0.501 $\pm$ 0.082 & 0.597 $\pm$ 0.093* & \textbf{0.650} $\pm$ \textbf{0.085}* \\
        \bottomrule
    \end{tabular}
\end{table}

On pseudo low-resolution scans (Table~\ref{tab:dice_scores_pseudo_singlecol}), segmentation performance improved significantly with SR methods, especially TVSRN-V2, which restored high accuracy and outperformed the other techniques.

\begin{table}[t]
    \centering
    \scriptsize  
    \caption{Dice scores for V-Net MTL on real low-resolution CT after SR. Bold indicates best. * denotes $p<0.05$ vs. Bicubic.}
    \label{tab:dice_scores_real_singlecol}
    \begin{tabular}{lccc}
        \toprule
        Class & Bicubic & TVSRN & TVSRN-V2 \\
        \midrule
        LR lobe & 0.897 $\pm$ 0.031 & 0.945 $\pm$ 0.028* & \textbf{0.961} $\pm$ \textbf{0.022}* \\
        MR lobe & 0.891 $\pm$ 0.038 & 0.919 $\pm$ 0.035* & \textbf{0.935} $\pm$ \textbf{0.030}* \\
        UR lobe & 0.902 $\pm$ 0.028 & 0.948 $\pm$ 0.031* & \textbf{0.966} $\pm$ \textbf{0.018}* \\
        LL lobe & 0.904 $\pm$ 0.032 & 0.944 $\pm$ 0.028* & \textbf{0.958} $\pm$ \textbf{0.025}* \\
        UL lobe & 0.895 $\pm$ 0.035 & 0.954 $\pm$ 0.031* & \textbf{0.965} $\pm$ \textbf{0.021}* \\
        Trachea$^{b}$ & 0.916 $\pm$ 0.029 & 0.954 $\pm$ 0.025* & \textbf{0.967} $\pm$ \textbf{0.019}* \\
        Bronchi$^{b}$ & 0.508 $\pm$ 0.076 & 0.608 $\pm$ 0.076* & \textbf{0.670} $\pm$ \textbf{0.070}* \\
        \bottomrule
    \end{tabular}
\end{table}

Lastly, on real-world low-resolution CT (Table~\ref{tab:dice_scores_real_singlecol}), TVSRN-V2 again outperformed all competing methods. These findings demonstrate that TVSRN-V2 not only enhances image quality but also substantially boosts segmentation performance across different CT conditions. This establishes it as a superior SR approach for robust downstream analysis in clinical workflows.

\subsection{Evaluating SR for Histology Classification and Prognosis in NSCLC}

This section evaluates the effectiveness of TVSRN-V2 as a preprocessing step for two key clinical tasks in non-small cell lung cancer (NSCLC): histology classification and patient prognosis. We focused on feature sets derived from radiomics and TMR-CT while comparing performance with and without SR using TVSRN-V2.

\subsubsection{Histology Classification Results}

As shown in Table~\ref{tab:SR_histology}, SR significantly improved F1-scores for adenocarcinoma (AC) vs. squamous cell carcinoma (SCC) classification across all datasets. The most notable gains were observed when using TMR-CT features, especially in datasets with thicker slices (e.g., ICHT and GSTT). For instance, in the ICHT dataset, the F1-score improved from 0.79$\pm$0.03 to 0.86$\pm$0.02 after applying SR. Improvements were statistically significant for TMR-CT in all datasets ($\emph{p} \leq 0.05$).

\begin{table*}[t]
    \centering
    \scriptsize
    \caption{Impact of SR on Histology Classification of NSCLC. F1-score ($\pm$ standard error) for classification of AC and SCC using Random Forest and selected feature sets. P-values from one-sided Wilcoxon signed-rank tests compare performance with and without SR.}
    \label{tab:SR_histology}
    \begin{tabular}{lcccccccccc}
        \toprule
        & & \multicolumn{2}{c}{TCIA (ext)} & \multicolumn{2}{c}{RMH} & \multicolumn{2}{c}{GSTT} & \multicolumn{2}{c}{ICHT} \\
        \cmidrule(lr){3-4} \cmidrule(lr){5-6} \cmidrule(lr){7-8} \cmidrule(lr){9-10}
        & & F1-score & P-value & F1-score & P-value & F1-score & P-value & F1-score & P-value \\
        \midrule
        \multirow{2}{*}{Radiomics}
        & w/o SR & 0.63$\pm$0.02 &  & 0.58$\pm$0.04 &  & 0.59$\pm$0.03 &  & 0.57$\pm$0.02 & \\
        & with SR & 0.65$\pm$0.02 & 0.12 & 0.59$\pm$0.04 & 0.32 & 0.60$\pm$0.03 & 0.22 & 0.60$\pm$0.03 & 0.34 \\
        \addlinespace
        \multirow{2}{*}{TMR-CT}
        & w/o SR & 0.84$\pm$0.03 &  & 0.78$\pm$0.02 &  & 0.77$\pm$0.03 &  & 0.79$\pm$0.03 & \\
        & \textbf{with SR} & \textbf{0.87$\pm$0.02} & \textbf{0.03} & \textbf{0.83$\pm$0.02} & \textbf{0.03} & \textbf{0.82$\pm$0.03} & \textbf{0.04} & \textbf{0.86$\pm$0.02} & \textbf{0.02} \\
        \bottomrule
    \end{tabular}
\end{table*}

\subsubsection{Prognosis Results}

Table~\ref{tab:SR_prognosis} reports the impact of SR on the concordance index (C-index) for NSCLC prognosis using a Random Survival Forest. TMR-CT features demonstrated the highest improvements across datasets after applying SR. For example, the C-index in the TCIA validation set increased from 0.74 to 0.80 with SR. P-values indicate significant improvements in multiple settings, particularly for datasets with thicker slices.

\begin{table*}[t]
    \centering
    \footnotesize
    \caption{Impact of SR on Prognosis of NSCLC. C-index ($\pm$ standard error) for Random Survival Forest using TMR-CT and radiomics features. P-values from one-sided Z-tests compare with and without SR.}
    \label{tab:SR_prognosis}
    \begin{tabular}{lcccccccccc}
        \toprule
        & & \multicolumn{2}{c}{TCIA (ext)} & \multicolumn{2}{c}{RMH} & \multicolumn{2}{c}{GSTT} & \multicolumn{2}{c}{ICHT} \\
        \cmidrule(lr){3-4} \cmidrule(lr){5-6} \cmidrule(lr){7-8} \cmidrule(lr){9-10}
        & & C-index & P-value & C-index & P-value & C-index & P-value & C-index & P-value \\
        \midrule
        \multirow{2}{*}{Radiomics}
        & w/o SR & 0.62$\pm$0.04 & & 0.58$\pm$0.05 & & 0.61$\pm$0.04 & & 0.59$\pm$0.06 & \\
        & with SR & 0.66$\pm$0.03 & \textbf{0.03} & 0.62$\pm$0.03 & 0.09 & 0.64$\pm$0.05 & 0.05 & 0.63$\pm$0.04 & 0.03 \\
        \addlinespace
        \multirow{2}{*}{TMR-CT}
        & w/o SR & 0.74$\pm$0.03 & & 0.72$\pm$0.04 & & 0.71$\pm$0.05 & & 0.71$\pm$0.04 & \\
        & \textbf{with SR} & \textbf{0.80$\pm$0.04} & \textbf{0.02} & \textbf{0.76$\pm$0.04} & \textbf{0.11} & \textbf{0.75$\pm$0.05} & \textbf{0.08} & \textbf{0.75$\pm$0.05} & \textbf{0.07} \\
        \bottomrule
    \end{tabular}
\end{table*}

\section{Discussion}

TVSRN-V2 tackles a significant challenge in the field of CT imaging: producing high-resolution images without increasing radiation dose or requiring costly hardware modifications. By effectively training on both real and pseudo low-resolution scans, the model demonstrates strong generalization across varying slice thicknesses and diverse clinical scenarios.

Our evaluation surpassed traditional image quality metrics like PSNR and SSIM, as we explored the impact of super-resolution (SR) on clinically relevant downstream tasks, including segmentation, radiomics, classification, and prognosis. TVSRN-V2 consistently surpassed baseline methods—specifically bicubic interpolation and the original TVSRN—highlighting both quantitative and qualitative improvements. With a PSNR of \(39.16 \pm 1.84\) and SSIM of \(0.941 \pm 0.028\), the model showcased high fidelity in preserving anatomical structures, which is vital for various clinical applications.

Ablation studies offered important insights into the model's architecture. Removing the Through-Plane Attention Block (TAB) or substituting Swin Transformer Layers (STL2) with standard ViT encoders resulted in marked performance degradation. These findings reinforce the significance of asymmetric decoding and volumetric attention in effectively modeling through-plane dependencies.

In segmentation tasks, TVSRN-V2 yielded improved Dice scores across both real and pseudo low-resolution CTs, validating its role as a preprocessing step for V-Net MTL. Even when pseudo low-resolution inputs lacked anatomical detail, the SR process significantly aided in recovering structures critical for precise lobar boundary prediction. Notably, the model produced the highest Dice scores when comparing outputs from SR-enhanced low-resolution scans against high-resolution-derived masks, indicating its robustness in various imaging contexts.

The radiomics analysis, conducted using the RIDER dataset, further demonstrated enhanced feature reproducibility following SR. Higher SSIM values correlated with increased intraclass correlation, suggesting that SR can stabilize radiomic features across longitudinal or variable-resolution scans, which is crucial for consistent clinical decision-making.

While our findings are promising, a limitation of this study is the homogeneity of the training data, which was sourced from a single scanner. Although the model was evaluated across multiple downstream tasks and external test sets, generalizability could be enhanced through multi-scanner training or adaptive fine-tuning strategies. Importantly, unlike the implications of a single scanner dataset, the clinical relevance of our evaluations extends to diverse imaging conditions.

Additionally, the computational demands of TVSRN-V2’s transformer architecture may hinder rapid deployment in some clinical settings. Future work should focus on improving model efficiency to facilitate quicker implementations.

Interestingly, SR provided the most substantial benefits on the ICHT dataset, characterized by the coarsest slice thickness. This insight suggests that TVSRN-V2 may be particularly advantageous in real-world scenarios where acquiring thin-slice CT scans is not feasible, thereby enhancing accessibility to high-quality imaging in diverse healthcare environments.

\section{Conclusion}

In this work, we presented TVSRN-V2, a transformer-based volumetric super-resolution (SR) framework specifically designed to enhance the quality of low-dose lung CT scans while facilitating downstream clinical tasks. By integrating Through-plane Attention Blocks (TAB) and Swin Transformer V2 within a scalable encoder-decoder architecture, our model effectively reconstructs high-frequency anatomical details and adapts well across diverse scanner protocols through pseudo-low-resolution augmentation.

We demonstrated notable improvements across three critical downstream lung cancer analysis tasks—lobe segmentation, radiomics, and prognosis—all validated on multi-cohort clinical datasets. The observed gains in segmentation accuracy, radiomic reproducibility, and prognostic performance underscore the clinical utility of transformer-based SR techniques in enabling more reliable and quantitative imaging assessments.

Overall, TVSRN-V2 presents a practical and robust solution for dose-efficient imaging and its potential for real-world deployment across heterogeneous healthcare settings. This contribution paves the way for more effective, data-driven decision support in thoracic oncology and signifies a step forward in enhancing patient care through advanced imaging technologies.

\bibliographystyle{plain} 
\bibliography{bib} 

\appendix

\section{Supplementary Material}

\section{Prognosis and Histology Dataset Details}

To evaluate the impact of SR on downstream clinical tasks, we used radiomics and TMR-CT features to assess histology classification and prognosis prediction across a multi-cohort NSCLC dataset. Specifically, we investigated whether SR preprocessing improves the performance and robustness of these features under varying imaging conditions.

The models were trained on the public TCIA dataset \cite{aerts2014decoding}, which includes 203 NSCLC patients (152 adenocarcinoma [AC], 51 squamous cell carcinoma [SCC]). A split of 120 patients for training/validation and 83 for internal testing was used, following prior work \cite{boubnovski2024deep}.

To evaluate generalizability and the effect of SR in heterogeneous clinical settings, we used three independent external test sets from the OCTAPUS-AI study (ClinicalTrials.gov ID: NCT04721444), comprising patients from Royal Marsden Hospital (RMH), Guy’s and St Thomas’ Hospital (GSTT), and Imperial College Healthcare NHS Trust (ICHT) \cite{hindocha2022gross}. These datasets vary significantly in image acquisition protocols, patient demographics, and slice thickness—ranging from 2–3 mm—providing an ideal benchmark for evaluating SR's effect under real-world conditions.

By comparing performance on original vs. SR-enhanced scans, we quantified improvements in classification (AC vs. SCC) and prognosis (C-index) using radiomic and TMR-CT features across all cohorts. A detailed breakdown of cohort characteristics, imaging parameters, and treatment distributions is available in Supplementary Table \ref{tab:institution_comparison}.

\begin{landscape}
\begin{table*}[htbp]
    \centering
    \scriptsize
    \caption{Comparison of patient characteristics across different institutions used to evaluate the impact of SR on NSCLC histology classification and prognosis.}
    \label{tab:institution_comparison}
    \begin{tabular}{llcccccccc|c}
        \toprule
        \textbf{Category} & \textbf{Variable} & \multicolumn{2}{c}{\textbf{TCIA (n=203)}} & \multicolumn{2}{c}{\textbf{GSTT (n=128)}} & \multicolumn{2}{c}{\textbf{ICHT (n=101)}} & \multicolumn{2}{c}{\textbf{RMH (n=310)}} & \textbf{p-value} \\
        \cmidrule(lr){3-4} \cmidrule(lr){5-6} \cmidrule(lr){7-8} \cmidrule(lr){9-10}
        & & AC & SCC & AC & SCC & AC & SCC & AC & SCC \\
        \midrule
        \textbf{Demographics} & Patients & 152 & 51 & 67 & 61 & 49 & 52 & 189 & 121 \\
        & Age (IQR) & 68($\pm$15) & 71($\pm$14) & 70($\pm$15) & 73($\pm$11) & 71($\pm$14) & 72($\pm$11) & 74($\pm$17) & 76($\pm$12) \\
        \midrule
        \multirow{2}{*}{\textbf{Gender}} & Male & 32 & 112 & 36 & 39 & 27 & 33 & 83 & 82 \multirow{2}{*}{$<$ 0.001} \\
        & Female & 120 & 40 & 31 & 22 & 22 & 19 & 106 & 39 \\
        \midrule
        \multirow{2}{*}{\textbf{CT type}} & Contrast & -- & -- & 29 & 23 & 29 & 31 & 55 & 42 \\
        & Non-contrast & -- & -- & 38 & 37 & 20 & 21 & 134 & 79 \\
        \midrule
        \textbf{Dosimetry} & BED (Gy) & -- & -- & 77($\pm$39) & 77($\pm$35) & 70($\pm$9) & 70($\pm$9) & 77($\pm$39) & 72($\pm$23) \\
        \midrule
        \multirow{2}{*}{\textbf{Outcomes}} & Survival days & 583 & 492 & 864 & 760 & 895 & 868 & 834 & 694 \\
        & Recorded deaths & 45 & 139 & 32 & 40 & 26 & 42 & 104 & 79 \\
        \midrule
        \multirow{4}{*}{\textbf{Treatment}} & RT only & -- & -- & 10 & 19 & 22 & 29 & 31 & 31 \\
        & SBRT & -- & -- & 29 & 17 & 0 & 0 & 90 & 31 \\
        & Sequential chemoRT & -- & -- & 10 & 11 & 9 & 9 & 37 & 37 \\
        & Concurrent chemoRT & -- & -- & 18 & 14 & 18 & 14 & 31 & 18 \\
        \midrule
        \multirow{3}{*}{\textbf{TNM8 Stage}} & I & -- & -- & 32 & 23 & 10 & 8 & 89 & 36 \\
        & II & -- & -- & 10 & 9 & 13 & 12 & 22 & 22 \\
        & III & -- & -- & 25 & 29 & 26 & 32 & 78 & 63 \\
        \midrule
        \multirow{3}{*}{\textbf{Slice Thickness (mm)}} & 2.0 & -- & -- & 0 & 0 & 0 & 0 & 172 & 114 \\
        & 2.5 & -- & -- & 67 & 61 & 0 & 0 & 17 & 7 \\
        & 3.0 & -- & -- & 0 & 0 & 49 & 52 & 0 & 0 \\
        \bottomrule
    \end{tabular}
\end{table*}
\end{landscape}

\section{ Lung Segmentation Dataset Details}
\subsection*{Dataset Description}
Our comprehensive dataset comprises 124 thoracic CT scans, including both normal and pathological cases. The dataset is structured into training (50 scans), validation (25 scans), test (25 scans), and external disease-specific test sets (24 scans).

\subsection{Training Dataset}
The training set consists of 50 'normal' or 'near normal' thoracic CT examinations (including cases with small nodules but preserved lung architecture). Thirty-eight cases were sourced from the SPIE-AAPM Lung CT Challenge cohort (TCIA dataset), with the remaining 12 from Imperial College Healthcare NHS Trust (REC: 18HH4616).

Images were acquired using:
\begin{itemize}
    \item SPIE-AAPM data: Philips Brilliance 64 scanner
        \begin{itemize}
            \item 1-mm section thickness, portal venous phase
            \item 120/140 kV tube potential
            \item 0.549-0.900 mm in-plane resolution (mean: 0.685 mm)
            \item "D" convolution kernel
        \end{itemize}
    \item Imperial College data: Philips Brilliance/Siemens Somatom Definition AS+
        \begin{itemize}
            \item 1-mm section thickness, portal venous phase
            \item 120 kV tube potential
            \item 0.625-0.750 mm in-plane resolution
            \item "I" convolution kernel
        \end{itemize}
\end{itemize}

\subsection{Validation and Test Datasets}
The validation and test sets (25 scans each) were derived from the LUNA16 competition dataset, selected from the LIDIC dataset (50.7\% female; median age = 60.1 years). These scans feature diverse technical parameters across various scanners, with section thickness <3 mm and consistent spacing.

\subsection{Disease-Specific External Test Set}
The external test set comprises four distinct pathological cohorts (6 scans each):
\begin{itemize}
    \item COVID-19 pneumonitis: Sourced from RICORD, PCR-positive with characteristic CT features
    \item Lung cancer: Cases showing significant architectural distortion
    \item Collapsed lung: Complete or significant partial single lobar collapse
    \item COPD: GOLD stages II-IV (mean emphysema: 13.1\%, SD: 3.3\%, defined as voxels $\leq$-950 HU)
\end{itemize}

\subsection{Segmentation Details}
All segmentations (pulmonary lobes, trachea, and bronchi) were prepared or validated by a board-certified radiologist with 4 years of chest imaging experience using:
\begin{itemize}
    \item Pulmonary Toolkit (MATLAB-based)
    \item 3D Slicer
\end{itemize}
Airway segmentations include lobar, segmental, and where possible, subsegmental airways, grouped by their respective lobes.

\end{document}